\documentclass[conference]{IEEEtran}
\ifCLASSINFOpdf
  \usepackage[pdftex,dvipdfm]{graphicx}
\else
\fi

\usepackage{amsmath,amssymb,amsfonts}
\usepackage{hyperref}
\usepackage{makecell}
\usepackage[most]{tcolorbox}
\usepackage{url}
\definecolor{mygreen}{RGB}{146,208,80}
\usepackage{listings}
\usepackage{ulem}
\usepackage{multirow, makecell}

\usepackage{bmpsize}

\newcommand{\quotebox}[1]
{
  \begin{center}
  \vspace{-10pt}
    \fcolorbox{white}{blue!15!gray!15}{
      \begin{minipage}{0.95\linewidth}
        \center
        \small
        \begin{minipage}{\linewidth}{\space\normalsize``}{\textit{#1}}{\normalsize\hfill''}
        \end{minipage}
      \end{minipage}
    }
\end{center}
}

\begin{document}
%
\title{Time for aCTIon: Automated Analysis of\\Cyber Threat Intelligence in the Wild}

\author{\IEEEauthorblockN{Giuseppe Siracusano\\Davide Sanvito\\Roberto Gonzalez}
\IEEEauthorblockA{\textit{NEC Laboratories Europe}}
\and
\IEEEauthorblockN{Manikantan Srinivasan\\Sivakaman Kamatchi}
\IEEEauthorblockA{\textit{NEC Corporation India}}
\and
\IEEEauthorblockN{Wataru Takahashi\\Masaru Kawakita\\Takahiro Kakumaru}
\IEEEauthorblockA{\textit{NEC}}
\and
\IEEEauthorblockN{Roberto Bifulco}
\IEEEauthorblockA{\textit{NEC Laboratories Europe}}}


%


\maketitle

\begin{abstract}
Cyber Threat Intelligence (CTI) plays a crucial role in assessing risks and enhancing security for organizations. However, the process of extracting relevant information from unstructured text sources can be expensive and time-consuming. Our empirical experience shows that existing tools for automated structured CTI extraction have performance limitations. Furthermore, the community lacks a common benchmark to quantitatively assess their performance. 

We fill these gaps providing a new large open benchmark dataset and aCTIon, a structured CTI information extraction tool.
The dataset includes 204 real-world publicly available reports and their corresponding structured CTI information in STIX format. Our team curated the dataset involving three independent groups of CTI analysts working over the course of several months. To the best of our knowledge, this dataset is two orders of magnitude larger than previously released open source datasets.
We then design aCTIon, leveraging recently introduced large language models (GPT3.5) in the context of two custom information extraction pipelines. We compare our method with 10 solutions presented in previous work, for which we develop our own implementations when open-source implementations were lacking.

Our results show that aCTIon outperforms previous work for structured CTI extraction with an improvement of the F1-score from 10\%points to 50\%points across all tasks. 
\end{abstract}


%

\section{Introduction}
Cyber Threat Intelligence (CTI) provides security operators with the information they need to protect against cyber threats and react to attacks~\cite{bromiley2016threat}. When structured in a standard format, such as STIX~\cite{barnum2012standardizing}, CTI can be used with automated tools and for efficient search and analysis~\cite{wagner2019cyber}. However, while many sources of CTI are structured and contain Indicators of Compromise (IoCs), such as block lists of IP addresses and malware signatures, most  CTI data is usually presented in an unstructured format, i.e., text reports and articles~\cite{hwang2022current}. This form of CTI proves to be the most helpful to security operators, since it includes information about the attackers (\textit{threat actors}) and victims (\textit{targets}), and how the attack is performed: tools (\textit{malwares}) and \textit{attack patterns}. Ultimately, this is the information that enables threat hunting activities~\cite{gao2021enabling, milajerdi2019poirot}.

Given the relevance of CTI, despite the limited resources, security analysts invest a significant amount of time to manually process sources of CTI to structure the information in a standard format~\cite{park2022full}. In fact, the effort is sufficiently large that companies form organizations to share the structured CTI and the cost of producing it. For instance, the Cyber Threat Alliance (CTA) provides a platform to share CTI among members in the form of STIX \textit{bundles}, and counts over 30 large companies among its members, such as CISCO, McAfee, Symantec, Sophos, Fortinet and others~\cite{cta}.  

To aid this activity, the security community has been actively researching ways to automate the process of extracting information from unstructured CTI sources, which led to the development of several methods and tools~\cite{husari2017ttpdrill,satvat2021extractor}. While these solutions contribute to reduce the analyst load, their focus has been historically limited to the extraction of IoCs, which are relatively easy to identify with pattern matching methods (e.g., regular expressions). Only recently, the advances in natural language processing (NLP) using deep learning have enabled the development of methods that can extract more complex information (i.e., threat actor, malware, target, attack pattern). Nonetheless, the performance of these solutions is still limited (Section~\ref{sec:eval}).

One of the problems is the way these machine learning solutions operate: they often specialize a general natural language processing machine learning model, fine-tuning it for the cybersecurity domain. Fine-tuning happens by means of providing the models with a training dataset, built by manually labeling a large number of reports. 
However, these AI models are specifically designed to perform tasks such as Named Entity Recognition (NER), which are close to the needs of a security analyst and yet crucially different. For instance, a report describing the use of a new malware might mention other known malwares in a general introductory section. These malwares would be extracted by a regular NER model, whereas a security analyst would ignore them when compiling the structured report. That is, generating a structured CTI report requires extracting only the \textit{relevant} named entities.
To make things worse, the security community currently lacks a large labeled dataset that could work as benchmark to evaluate these tools. Indeed, the current state-of-the-art is mostly evaluated using metrics belonging to the NLP domain, which essentially evaluate a subtask in place of the end-to-end task performed by the security analyst.

Our goal is to provide a means to evaluate existing and future tools for structured CTI information extraction, and a solution to improve on the state-of-the-art. 

First, we contribute a labeled dataset including 204 reports collected from renowned sources of CTI, and their corresponding STIX bundles. The reports vary in content and length, containing 2133 words on average and up to 6446. Our team of trained security analysts examined the reports over the course of several months to define the corresponding STIX bundles.
 This process requires, among other things, to classify \textit{attack patterns} using the MITRE ATT\&CK Matrix (tactics, techniques, and procedures), which includes more than 190 detailed entries \cite{mitre-attack}. The analyst needs to know these techniques and understand if the case described in the report fits any of them, to perform correct classification.

Second, we replicate the results of 10 recent works, providing our own implementations when these were not available, and use our benchmark dataset to evaluate them. Our evaluation shows that the improvement in NLP technology had a significant impact on the performance of the tools, which got much better over time and since the inclusion of NLP technology such as Transformer Neural Networks (e.g., BERT \cite{kenton2019bert}). At the same time, the evaluation shows there are still significant gaps, with the best performing tools achieving on average across all reports less than 50\% in recall/precision, for any specific type of information extracted (i.e., malware, threat actor, target and attack pattern).   

Finally, inspired by recent advances in Large Language Models (LLMs) such as GPT3~\cite{brown2020language}, we contribute a new solution, aCTIon, using LLM's zero-shot prompting and in-context learning capabilities.
Our approach addresses some of the main shortcomings and constraints of the current generation of LLMs, namely \textit{hallucinations}~\cite{lee2022factuality} and small context windows~\cite{stanford-long-learning}, in the constrained setting of our use case.
To do so, we introduce a novel two-step LLM querying procedure, which resembles some of the recent approaches used in the design of LLM-based generative AI Agents~\cite{yao2022react}. 
In the first step, we pre-process the input report to extract and condense information in a text that can fit the limits of the target LLM. In the second step, we define extraction and self-verification prompts for the LLM, which finally selects and classifies the extracted information.
We experiment with several alternative variations of the above general approach, and find that aCTIon can outperform the state-of-the-art by increasing the F1-score by 15-50\% points for malware, threat actor and target entities extraction, and by about 10\% points for attack pattern extraction.

To foster further research in this area, we release our dataset, including reports and labels.

\section{Life and pain of a CTI analyst}
\label{sec:background}

A large amount of valuable CTI is shared in unstructured formats, including open-source intelligence (OSINT), social media, the dark web, industry reports, news articles, government intelligence reports, and incident response reports. Using unstructured CTI is challenging as it cannot be efficiently stored, classified and analyzed, requiring security experts to thoroughly read and comprehend lengthy reports. Consequently, one of the tasks of a security analyst is to convert the vast amount of unstructured CTI information in a format that simplifies its further analysis and usage.

\begin{figure}[!t]
\centering{\includegraphics[width=0.9\columnwidth]{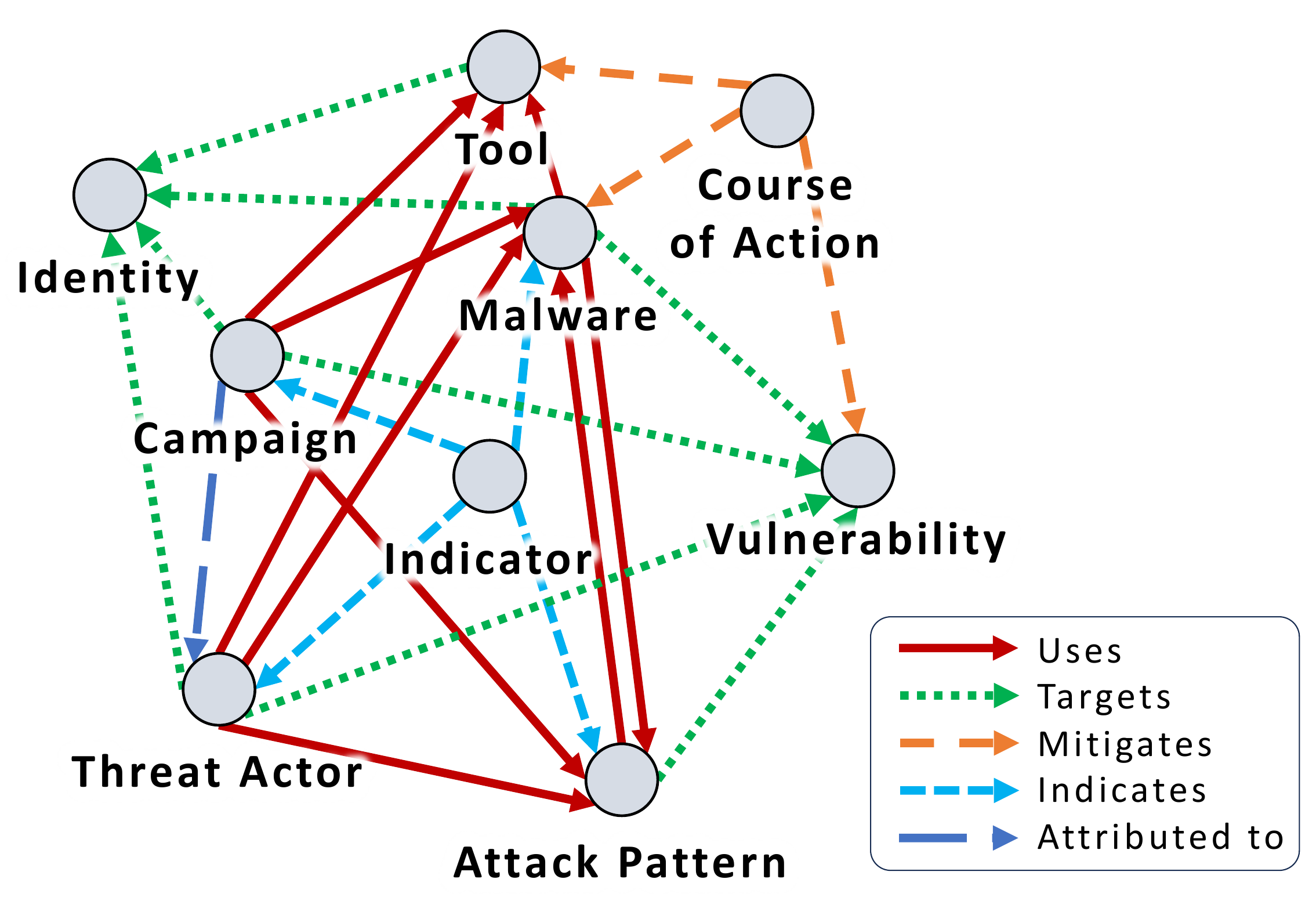}}
\caption{A subset of the STIX ontology, including all entities and relations contained at least once in our dataset.}
\label{fig:empirical_stix_ontology}
\end{figure}

STIX~\cite{barnum2012standardizing} is an example of a standard format for CTI widely adopted by the industry. In STIX, each \textit{report} (a \texttt{bundle} in STIX terminology) is a knowledge graph, i.e., a set of nodes and relations that describe a security incident or a relevant event. The STIX ontology describes all the entity and relation types: Figure \ref{fig:empirical_stix_ontology} shows a subset of the STIX ontology. 
The ontology includes several conceptual entities, such as \texttt{Threat Actor}, \texttt{Malware}, \texttt{Vulnerability}, \texttt{Attack Pattern}, and \texttt{Indicator}. Furthermore, it also defines relations between these entities, such as \texttt{uses} and \texttt{targets}, to capture their interactions.

In the remainder of this section we provide an example of a report, and introduce how analysts extract structured STIX bundles from text reports. We focus on the most common information extracted by analysts: 

\begin{itemize}
    \item Who performed the attack (i.e., \texttt{Threat Actor}), 
    \item Against whom it was performed (i.e. \texttt{Identity} pointed by a \texttt{targets} relation).
    \item How the attack was performed (i.e., \texttt{Malware} and \texttt{Attack Pattern}), 
\end{itemize}

This subset of the STIX's ontology is the most common set of information pieces contained in reports. For instance, in our dataset 75\% and 54\% of reports include at least a Malware and Threat Actor entity, respectively. Furthermore, and more importantly for a fair evaluation of the state-of-the-art, this subset is consistently supported across existing tools and previous work, which allows us to run an extensive comparison among solutions.

\subsection{Structured CTI Extraction}
\label{sec:our_task}

\begin{figure}[t]
\centering{\includegraphics[width=1\columnwidth]{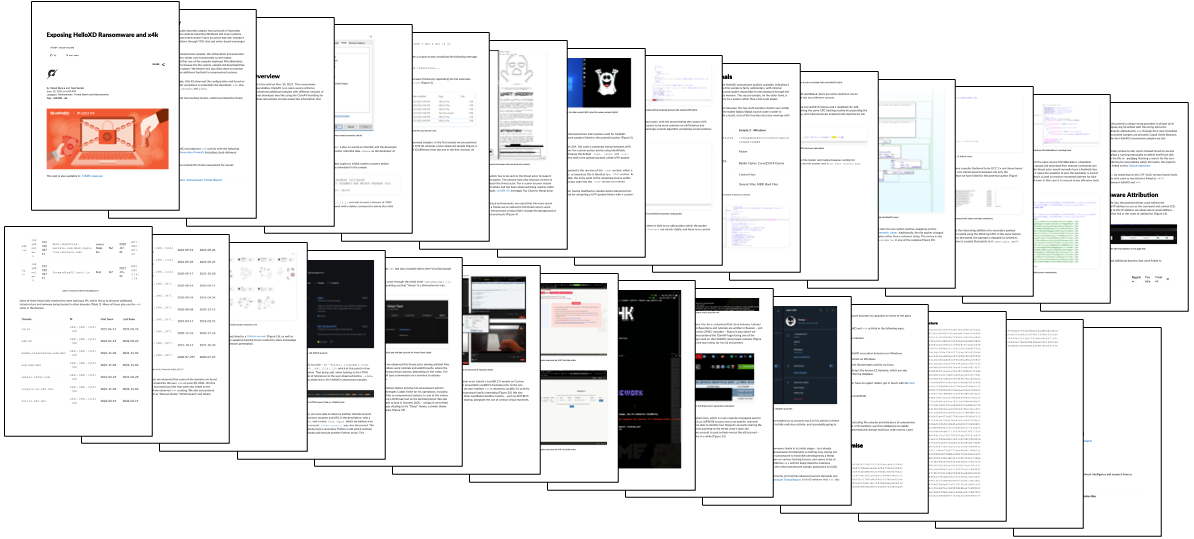}}
\caption{An example of a report published by Palo Alto Networks\textsuperscript{\ref{fn:report_url}}. While Indicators of Compromise are easy to extract being collected at the end of the report, extracting \texttt{Threat Actor}, \texttt{Malware}, \texttt{Attack Pattern} and the other STIX's entities requires security experts to perform manual analysis.}
\label{fig:report}
\end{figure}

\begin{figure}[]
\centering{\includegraphics[width=0.9\columnwidth]{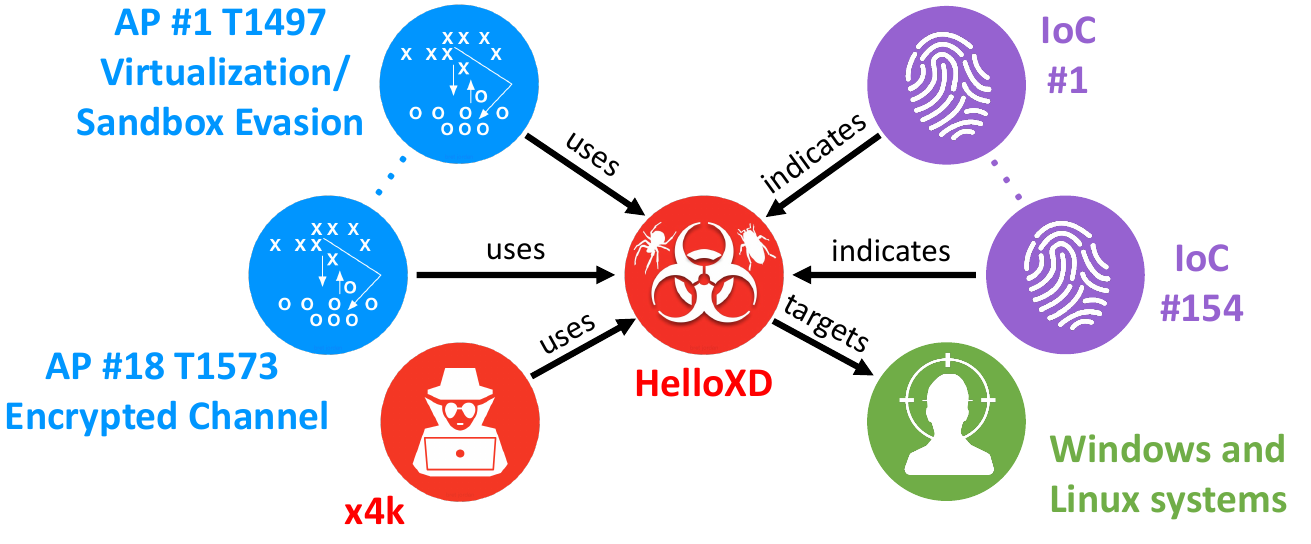}}
\caption{A STIX bundle describing the report from Figure~\ref{fig:report}.}
\label{fig:bundle}
\end{figure}

To illustrate the \textit{structured CTI extraction} task, we consider
a technical blog post from Palo Alto Networks\footnote{\label{fn:report_url} \url{https://unit42.paloaltonetworks.com/helloxd-ransomware/}}, presented at a glance in Figure~\ref{fig:report}. The report describes the attribution of the ransomware \textit{HelloXD} to a the threat actor known as \textit{x4k}, including the set of tactics, techniques and procedures associated with them. 
The report is about 3.7K words long, includes 24 different images, 3 tables with different information and a list of Indicator of Compromise (in a dedicated section at the end of the report). It first explains the functionality of the \textit{HelloXD} ransomware, and then it uncovers several clues that link the ransomware to the threat actor \textit{x4k}. 
Furthermore, the post provides a description of the threat actor's modus operandi and infrastructure.

Structured CTI extraction is to define a STIX \texttt{bundle} representing the report, like the one depicted in Figure~\ref{fig:bundle}. The bundle includes the \texttt{Threat Actor} (x4k), the \texttt{Malware} (HelloXD), and a set of \texttt{Attack Pattern} entities describing the various tactics, techniques and procedures, plus the \texttt{Indicator}s extracted from the last section of the report.

Defining this bundle is a time consuming task that requires security knowledge and experience. It can take 3-10 hours to extract a structured STIX bundle out of a report. For instance, in \cite{park2022full} the authors mention that labelling 133 reports required 3 full time annotators over 5 months. Likewise, the annotation of the 204 reports in the dataset we release with this paper took our team of CTI analysts several months.

To understand why this task is time consuming, and why it is hard to automate, let us consider how analysts identify the relevant entities and the case of our sample report.

\vspace{0.2cm}
\noindent\textbf{Malware, Threat Actor, Identity}
First, the analyst typically starts identifying malwares, threat actors and identities. While this might at first glance appear as a simple task, security reports tend to be semantically complex, including issues such as: information represented ambiguously, e.g., threat actor and malware called with the same name; the use of aliases to describe the same entity, e.g., a malware called with multiple name variants; uncertain attribution of attacks, i.e., the report might certainly attribute some attacks to a threat actor, and only mention other attacks as potentially related to the threat actor, but not confirmed. These are just few examples of the nuances that make processing time consuming, and automation difficult.

\noindent\textit{Example:} our sample report specifically discusses the \textit{HelloXD} ransomware. Yet, it is not uncommon for a malware to be deployed and utilized in conjunction with other malicious software. 
Thus, understanding what malicious software is effectively described in the attack is critical to understanding which malware nodes should be included in the STIX bundle. 
In the report, there are mentions of two other malwares beyond \textit{HelloXD}: \textit{LockBit 2.0} and \textit{Babuk/Babyk}. 
However, these malwares should NOT be included in the bundle. 
For instance, \textit{LockBit 2.0} is mentioned because it leverages the same communication mean used by \textit{HelloXD} (see quote below). Nonetheless, \textit{LockBit 2.0} is not directly connected to \textit{HelloXD} infections, and therefore it should not be included.
\quotebox{
The ransom note also instructs victims to download Tox and provides a Tox Chat ID to reach the threat actor. Tox is a peer-to-peer instant messaging protocol that offers end-to-end encryption and has been observed being used by other ransomware groups for negotiations. For example, LockBit 2.0 leverages Tox Chat for threat actor communications.
}

\vspace{0.2cm}
\noindent\textbf{Attack Pattern} Second, the analyst has to identify the attack patterns, i.e., descriptions of tactics, techniques and procedures, and attribute them to the  entities identified in the previous step. This introduces additional challenges: attack patterns are \textit{behaviors} typically described throughout several paragraphs of the report, and they are collected and classified in standard taxonomies such as the MITRE ATT\&CK Matrix~\cite{mitre-attack}. The MITRE ATT\&CK Matrix includes more than 190 detailed techniques and 400 sub-techniques. The analyst has to refer to them when building the bundle, identifying which of the classified techniques are contained in the text report.
That is, this tasks requires both understanding of the report and extensive specialized domain knowledge.

The following is an example of how an Attack Pattern is described in a text report\footnote{\url{https://www.proofpoint.com/us/blog/threat-insight/good-bad-and-web-bug-ta416-increases-operational-tempo-against-european}}:

\quotebox{
TA416 has updated the payload by changing both its encoding method and expanding the payloads configuration capabilities
}

The analyst has first to identify the sentence containing the attack pattern, and then map it to the corresponding MITRE definition, that in the case of the example is the technique T1027 \textit{"Obfuscated Files or Information"}\footnote{\url{https://attack.mitre.org/techniques/T1027/}} described as:

\quotebox{
Adversaries may attempt to make an executable or file difficult to discover or analyze by encrypting, encoding, or otherwise obfuscating its contents on the system or in transit.
}

\noindent\textbf{Relevance}
Throughout the process, the analyst has to take decisions about what to leave out of the bundle, using their experience. This decision usually includes considerations about the level of confidence and details of the information described in the report. 
For instance, our sample report describes other activities related to the threat actor \textit{x4k}, such as the deployment of Cobalt Strike Beacon and the development of custom Kali Linux distributions. The analyst must determine whether to include this information or not. In this example, these other activities are just mentioned, but they are not related to the main topic of the report (nor contain enough details), and therefore they should not be included. 

\subsection{The quest for automation}
\label{sec:background:automation}
Given the complexity of the task, several solutions have been proposed over time to help automate structured CTI extraction~\cite{park2022full,satyapanich2020casie,husari2017ttpdrill,alam2022looking,orbinato2022automatic,gao2022threatkg,legoy2020automated}. Previous work addressed either individual problems, such as attack patterns extraction, or the automation of the entire task. Nonetheless, all the proposed tools still require significant manual work in practice. We find evidence for this claim in the empirical work of our team of CTI analysts, and we also confirm this later in the paper when introducing our evaluation (Section~\ref{sec:eval}).

We speculate one of the reasons for which existing solutions do not meet the expectations of CTI analysts is due to the lack of a benchmark that correctly represents the structured CTI extraction task, with its nuances and complexities. In particular, since previous work heavily relies on machine learning methods for natural language processing (NLP), it is quite common to resort to typical NLP evaluation approaches. However, NLP tasks are a conceptual subset of the end-to-end structured CTI extraction task, therefore, the majority of proposed benchmarks do not evaluate \textit{CTI-metrics}.

To exemplify this issue, let us consider Named Entity Recognition (NER), i.e., the NLP task of automatically identifying and categorizing relevant entities in a text.
When evaluating a NER component, NLP-metrics count how many times a word representing an entity is identified. For instance, if the malware \textit{HelloXD} is mentioned and identified 10 times, it would be considered as 10 independent correct samples by a regular NER evaluation. We refer to this approach as \textit{word-level}\footnote{For the sake of simplicity of exposition we use the term "word-level" in place of the more appropriate "token-level".} labeling. However, for structured CTI extraction, our interest is to extract the malware entity, regardless of how many times it appears in the report. This can potentially lead to an overestimation of the method performance. More subtly, as we have seen with the example of the Lockbit 2.0 malware, some entities that would be correctly identified by a NER tool are not necessarily relevant for CTI information extraction task. However, such entities are typically counted as correct if the evaluation employs regular NLP metrics. The same issue applies to the more complex Attack Pattern extraction methods. Indeed, they are commonly evaluated on sentence classification tasks, which assess the method ability to recognize if a given sentence is an attack pattern and to assign it to the correct class. We refer to this approach as \textit{sentence-level} labeling.
However, such metrics do not fully capture the performance of the method in the light of the CTI-metrics. This would involve identifying all \textit{relevant} attack patterns in a given report, and correctly attributing them to the relevant entities.

\begin{table}[]
\caption{Manually annotated reports. 
In some works, the annotated reports are a subset of a much larger dataset, whose size is reported in parenthesis. Numbers with a "*" refer to sentences rather than whole reports. (\checkmark) refers to datasets only partially released as open-source.}
\begin{center}
\small
\begin{tabular}{|l||c|c|c|c|}
\hline
 & \makecell{ Entities \& \\ relations } & \makecell{ Attack \\ patterns } & \makecell{ CTI \\ metrics }  & Public  \\ \hline \hline
SecIE    & 133 & 133 & - & \\ \hline
CASIE    & 1k & - & - & \checkmark \\ \hline
ThreatKG & 141 (149k) & 141 (149k) & - & \\ \hline
LADDER   & 150 (12k) & 150 (12k) & 5 & (\checkmark) \\ \hline
SecBERT  & - & 12.9k* & 6 & \checkmark \\ \hline
TRAM     & - & 1.5k* & - & \checkmark \\ \hline
TTPDrill & - & 80 (17k) & 80 & \\ \hline
AttacKG  & - & 16 (1.5k) & 16 & \\ \hline
rcATT    & - & 1.5k & - & \checkmark  \\ \hline
\end{tabular}
\end{center}
\label{table:e2e_datasets}
\end{table}

Table \ref{table:e2e_datasets} summarizes datasets from the literature, which are employed in the evaluation of the respective previous works. The table considers separately the extraction of the Attack Pattern entity, given its more complex nature compared to other entities (e.g., Malware, Threat Actor, Identity).
Some of these works use remarkably large datasets to evaluate the information extracted in terms of NLP performance (e.g., number of extracted entities). Unfortunately, they often include much smaller data subsets to validate the methods in respect to CTI-metrics. One reason often mentioned for the small size of such data subsets is the inherent cost of performing manual annotation by expert CTI analysts.

\noindent\textbf{NLP-metrics} SecIE \cite{park2022full}, ThreatKG \cite{gao2022threatkg} and LADDER \cite{alam2022looking}, all adopt datasets with word-level or sentence-level labeling.
Similarly, CASIE \cite{satyapanich2020casie} provides a large word-level labeled dataset, and does not cover attack patterns.
TRAM \cite{tram-github} and rcATT \cite{legoy2020automated} only provide attack pattern datasets that are sentence-level labeled.

\noindent\textbf{CTI-metrics} Only a few works provide labeled data that correctly capture CTI-metrics.
TTPDrill \cite{husari2017ttpdrill} and AttacKG \cite{li2022attackg} perform the manual labeling of 80 and 16 reports, respectively, on a reports-basis, but unfortunately do not share them. Also, they cover only attack patterns extraction.
SecBERT \cite{orbinato2022automatic} evaluates the performance on a large sentence-level dataset, but then provides only 6 reports with CTI-metrics.
Similarly, as part of its evaluation LADDER \cite{alam2022looking} also includes the Attack Pattern Extraction task using CTI-metrics, but just on 5 reports (which are not publicly shared).

\section{The Dataset}

\label{sec:dataset}
The lack of an open and sufficiently large dataset focused on structured CTI extraction hinders our ability to evaluate existing solutions and to consistently improve on the state-of-the-art. 
To fill this gap we created a new large dataset including 204 reports spanning 12 months since February 2022, and their corresponding STIX bundles, as extracted by our expert team of CTI analysts. The dataset represents real-world CTI data, as processed by security experts, and therefore exclusively focuses on CTI-metrics. 
We make the dataset publicly accessible.\footnote{The link to the repository is hidden for anonymity.}

The remainder of this section provides information about the dataset creation methodology, and introduces high-level statistics about the data.

\subsection{Methodology}
\label{sec:dataset:methodology}
Our organization includes a dedicated team of CTI analysts whose main task is to perform structured CTI extraction from publicly available sources of CTI. 
We leverage their expertise and established methodology to create a dataset of unstructured reports, and their corresponding extracted STIX bundles.
Among the reports daily processed by our CTI analysts team we selected a subset of 204 publicly available reports published by well-known relevant sources (cf. Section \ref{sec:dataset:summary}) and we further manually verified the classification of each of them.
Structured CTI extraction is manually performed by CTI analysts, organized in three independent groups with different responsibilities, as outlined next:

\hfill\\\textbf{Group A} selects unstructured reports or sources of information for structured CTI extraction. The selection is based on the analyst's expertise, and is often informed by observed global trends. This group is formed by a variable number of people, usually from two to four.
\hfill\\\textbf{Group B} performs a first pass of structured CTI extraction from the selected sources. This group makes large use of existing tools to simplify and automate information extraction. Notice that this set of tools partially overlaps with those we mentioned in Table~\ref{table:e2e_datasets} and that we later assess in our evaluation in Section~\ref{sec:eval}. The actual structured CTI extraction happens in multiple processing steps. First, the report is processed with automated parsers\footnote{Ad-hoc parsers are developed by the CTI analysts team whenever a new web source is consulted.}, e.g., to extract text from a web source. Second, the retrieved text is segmented into groups of sentences. These sentences are then manually analyzed by the analyst, who might further split, join, or delete sentences to properly capture concepts and/or remove noise. The final result is a set of paragraphs. Third, the analyst applies automated tools to do a pre-labeling of the named entities, and then performs a manual analysis on each single paragraph flagging the entities that are considered relevant and correct, potentially adding entities that were not detected. At this stage, the use of automated entity extraction tools expedites the tagging process for analysts. Named entities within the text are highlighted, accelerating the reading process, however, the accuracy of automatically assigned labels will undergo further verification to ensure their correctness.
Fourth, a second analysis is performed on the same set of paragraphs, this time to extract attack patterns. 
Also in this case, automated tools are used in the analysis process. The Logistic Regression model implemented in TRAM~\cite{tram-github} is used to identify sentences that clearly contain attack patterns.
The classification of these identified sentences can be quickly verified. Subsequently, the remaining text, which does not contain clear or explicit descriptions of attack patterns, is manually analyzed and classified.
By initially identifying obvious attack pattern definitions in the text, not only the amount of text requiring in-depth analysis is reduced, but it also simplifies the classification of ambiguous sentences. 
This group is composed of two analysts, with each analyst processing different reports.
Finally, the analyst uses a visual STIX bundle editor (an internally built GUI) to verify the bundle and check any correct attribution, i.e., the definition of relations among entities.
\hfill\\\textbf{Group C} performs an independent review of the work performed by Group B. 
The review includes manual inspection of the single steps performed by Group B, with the goal of accepting or rejecting the extracted STIX bundle.
This group is also composed of two analysts. The members of Group C and Group B switch roles for each analyzed report.

The above process is further helped by a software infrastructure that our organization developed specifically to ease the manual structured CTI extraction tasks. Analysts connect to a web application and are provided with a convenient pre-established pipeline of tools. Furthermore, the web application keeps track of their interactions and role (e.g., analyst vs reviewer), and additionally tracks the time spent for each sub-step of the process. This allows us to estimate the time spent to perform structured CTI extraction on a report (excluding the work of Group A). For the dataset presented in this paper, we observed an average of 4.5h per report, with the majority of the time spent by group B (about 3h).

In addition to the process described above, we incorporated an additional step involving a group of two researchers. This step focused on validating a subset of reports that were processed by Groups B and C. Firstly, we selected reports that were publicly available on the Internet and published by well-known organizations and institutions. Next, the two researchers independently relabeled and classified these reports. As a result of this step, we obtained a final dataset consisting of 204 reports where Groups B, C, and the researchers unanimously agreed on the labels and classifications.
During this validation process, the group of researchers directly consulted the original web sources to avoid any potential, even if remote, errors introduced by the automated web parser used by Group B.

\subsection{Dataset Summary}

\label{sec:dataset:summary}

\begin{table}[]
\caption{Dataset statistics: numer of reports from each source, and number of words and sentences in each report.}
\begin{center}
\small
\begin{tabular}{|c||c|c|c|c|}
\hline
                & Min & Avg & 95p & Max \\ \hline \hline
    reports/source & 1 & 3.3 & 9 & 11 \\ \hline
    words/report &  504 & 2133.6 & 4015.8 &  6446 \\ \hline
    sentences/report  &  11 &  86.3 & 172.5 &   358 \\ \hline

\end{tabular}
\end{center}
\label{table:dataset_stats}
\end{table}

The final dataset comprises 204 reports and their corresponding STIX bundle representations. The reports are published by renowned organizations and institutions such as Palo Alto Networks, Trend Micro, and Fortinet, resulting in a total of 62 different sources, with each source providing on average 3.3 reports (cf. Table \ref{table:dataset_stats}). 
Approximately, 79\% of our sources are also referenced on the official MITRE ATT\&CK website as external references when providing procedure examples for specific attack pattern techniques. 
This confirms that the selected reports are representative of a wide well-known body of CTI sources. 

\begin{table}[]
\caption{Topics covered by the reports in the dataset.}
\begin{center}
\small
\begin{tabular}{|l||ll|}
\hline
\multicolumn{1}{|c||}{Topic}           & \multicolumn{2}{c|}{Quota}                        \\ \hline \hline
\multicolumn{1}{|l||}{Malware}             & \multicolumn{1}{l|}{30\%} & \multirow{4}{*}{75\%} \\ \cline{1-2}
\multicolumn{1}{|l||}{Malware + Threat Actor}        & \multicolumn{1}{l|}{30\%} &                       \\ \cline{1-2}
\multicolumn{1}{|l||}{Malware + Threat Actor + Vulnerability} & \multicolumn{1}{l|}{8\%}  &                       \\ \cline{1-2}
\multicolumn{1}{|l||}{Malware + Vulnerability}      & \multicolumn{1}{l|}{7\%}  &                       \\ \hline
\multicolumn{1}{|l||}{Threat Actor}              & \multicolumn{1}{l|}{11\%} & \multirow{2}{*}{15\%} \\ \cline{1-2}
\multicolumn{1}{|l||}{Threat Actor + Vulnerability}       & \multicolumn{1}{l|}{4\%}  &                       \\ \hline
\multicolumn{1}{|l||}{Others}           & \multicolumn{2}{c|}{10\%}                        \\ \hline
\end{tabular}
\end{center}
\label{table:dataset_topics_stats}
\end{table}

Table \ref{table:dataset_topics_stats} presents the main topics covered in the reports. Approximately 75\% of the reports focus on a specific malware, often accompanied by information about the threat actor utilizing or developing the malware (around 30\%). 
Additionally, around 7\% of the reports discuss related vulnerabilities, while approximately 8\% provide details about both the associated threat actor and the exploited vulnerability. 
Some reports do not include malware entities but specifically describe threat actors (approximately 11\%) or vulnerabilities associated with them (around 4\%). 
The remaining 10\% of reports cover topics such as attack campaigns and vulnerabilities.

The selected reports in the dataset encompass approximately 90\% of the attack pattern classes found in the MITRE ATT\&CK Matrix for Enterprise. 
In addition, the dataset covers all the 10 most prevalent MITRE ATT\&CK tactics and techniques leveraged by attackers in 2022 \cite{picuslabs2023} with tens of reports mentioning each one of them. In total, the reports mention 188 different malware variants and 91 different threat actors.

The resulting dataset comprises 204 STIX bundles, which collectively contain 36.1k entities and 13.6k relations. Figure \ref{fig:empirical_stix_ontology} presents the resulting STIX ontology based on our dataset, including 9 unique entity types and 5 unique relation types. The figure also shows the set of admissible types of relations between specific pairs of entity types.

\begin{table}[]
\caption{Dataset statistics by STIX bundle. The last column shows the quota of bundles containing each type of entity or relation at least once.}
\begin{center}
\small
\begin{tabular}{|c||c|c|c|c|c|}
\hline
                & Min & Avg & 95p & Max & Quota \\ \hline
\hline
    STIX objects  &  13 & 177.1 & 525.8 &  1255 & - \\ \hline
  STIX relations  &   5 &  67.0 & 180.3 &   429 & - \\ \hline
\hline
             Malware & 0 &   0.9 &   2.0 &     5 & 75\% \\ \hline
        Threat Actor & 0 &   0.6 &   1.0 &     2 & 54\% \\ \hline
      Attack Pattern & 0 &  21.8 &  40.0 &    63 & 99\% \\ \hline
            Identity & 1 &   1.7 &   2.0 &     5 & 100\% \\ \hline
           Indicator & 1 &  41.9 & 163.1 &   395 & 100\% \\ \hline
            Campaign & 0 &   0.6 &   1.0 &     4 & 55\% \\ \hline
       Vulnerability & 0 &   0.5 &   2.0 &    11 & 21\% \\ \hline
                Tool & 0 &   0.1 &   1.0 &    10 & 6\% \\ \hline
    Course of Action & 0 &   0.0 &   0.0 &     1 & 2\% \\ \hline
\hline
                uses & 1 &  23.6 &  48.8 &    64 & 100\% \\ \hline
           indicates & 1 &  41.9 & 163.1 &   395 & 100\% \\ \hline
             targets & 0 &   1.2 &   3.8 &    12 & 77\% \\ \hline
       attributed-to & 0 &   0.3 &   1.0 &     2 & 26\% \\ \hline
           mitigates & 0 &   0.0 &   0.0 &     2 & 2\% \\ \hline
\end{tabular}
\end{center}
\label{table:dataset_stix_stats}
\end{table}

Table \ref{table:dataset_stix_stats} reports the dataset statistics of the STIX bundles associated to the reports and is split in three sections: total number of STIX objects and relations, number of STIX objects by type and number of STIX relations by type. For the last two sections, the last column provides the quota of bundles that includes at least once a given type of entity and relation.
For example, 75\% of the bundles include a Malware entity, and 54\% include a Threat Actor. This highlights the prevalence of these critical components within the dataset, underscoring their importance in the context of CTI extraction and analysis.

\section{It is time for \MakeLowercase{a}CTI\MakeLowercase{on}}
\label{sec:action}

The dataset introduced in Section~\ref{sec:dataset} allows us to quantify the performance of existing tools for structured CTI extraction. We presents those results in details later in Section~\ref{sec:eval}, but anticipate here that our empirical experience is confirmed: the performance of previous work on structured CTI extraction is still limited. For example, the best performing tools in the state-of-the-art provide at most 60\% F1-score when extracting entities such as \texttt{Threat Actor} or \texttt{Attack Pattern}.

Given the pressure to reduce the cost of structured CTI extraction in our organization, the limitations of the state-of-the-art, and the recent emergence of a new wave of powerful machine learning technologies for natural language processing, such as Large Language Models (LLM), we developed aCTIon: a structured CTI extraction framework. Our goal is to significantly simplify, or ideally entirely replace, the information extraction step from the task, i.e., the work of Group B described in Section~\ref{sec:dataset:methodology}, focusing most of the manual effort only on the bundle review step (the work of Group C).

aCTIon builds on the recent wave of powerful LLMs, therefore we provide first a short background about this technology, before detailing aCTIon's design goals and decisions.

\subsection{Large Language Models primer}
LLMs~\cite{bommasani2021opportunities} are a family of neural network models for text processing, generally based on Transformer neural networks~\cite{vaswani2017attention}. Unlike past language models trained on task-specific labeled datasets, LLMs are trained using unsupervised learning on massive amount of data. While their training objective is to predict the next word, given an input prompt, the scale of the model combined with the massive amount of ingested data makes them capable of solving a number of previously unseen tasks and acquire emergent behaviors~\cite{brown2020language}. For instance, LLMs are capable of translating from/to multiple languages, perform data parsing and extraction, classification, summarization, etc. 
More surprisingly, these emergent abilities include creative language generation, reasoning and problem-solving, and domain adaptation~\cite{wei2022emergent}. 

From the perspective of system builders, perhaps the most interesting emergent ability of LLMs is their \textit{in-context learning} and instruction-following capabilities. That is, users can \textit{program} the behavior of an LLM prompting it with specific natural language instructions. This removes the need to collect specific training data on a per-task basis, and enables their flexible inclusion in system design. 
For instance, a prompt like "Summarize the following text" is sufficient to generate high-quality summaries.

While LLMs have great potential, they also have significant limitations. First, their training is very expensive, therefore they are retrained with low frequency. This makes the LLM unable to keep up to date with recent knowledge. Second, their prompt input and output sizes are generally limited. The input of LLMs is first tokenized, and then provided to the LLM. A \textit{token} can be thought as a part of a word. For instance, a model might limit to 4k tokens (about 3k-3.5k words) the total size of input plus output, which limits the kind of inputs that can be processed. Finally, LLMs might generate incorrect outputs, a phenomenon sometimes called \textit{hallucination}~\cite{lee2022factuality}. In such cases, the LLM generated answers might be imprecise, incorrect, or even completely made-up, despite appearing as a confident statement at first glance.

\subsection{aCTIon - Design}
\label{sec:action:design}
\begin{figure}[!t]
\centering{\includegraphics[width=1\columnwidth]{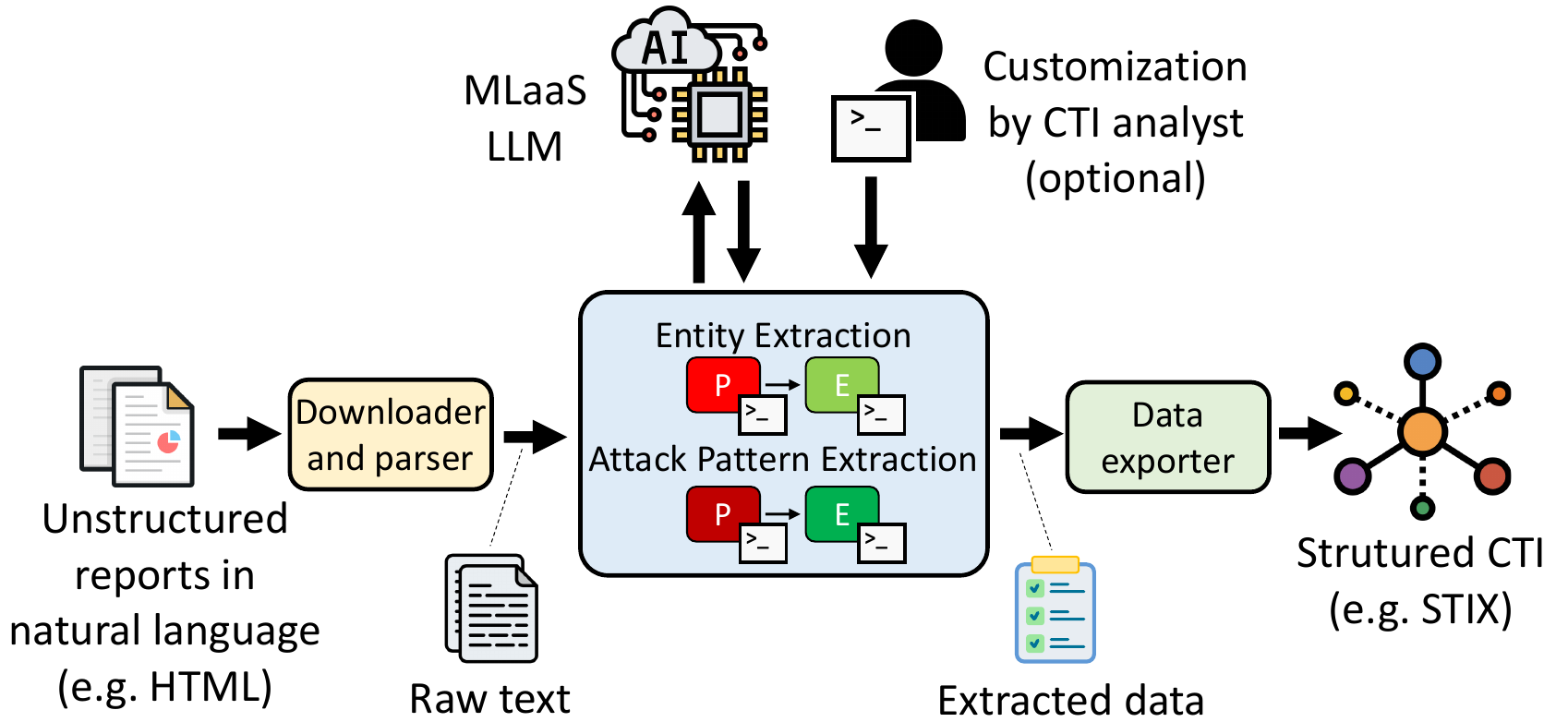}}
\caption{aCTIon high-level architecture}
\label{fig:action_architecture}
\end{figure}

Figure \ref{fig:action_architecture} depicts the overall architecture of the aCTIon framework, which comprises three main components. \textit{Downloader and parser} converts unstructured input reports in different formats, e.g., HTML, to text-only representations. It includes plugins to handle the format of specific well-known CTI sources, as well as a fallback default mode if no plugin is available for the desired source.
The second component is the core of the aCTIon framework and consists of two different pipelines: a pipeline extracts most of the entities and relations; a second pipeline deals specifically with attack pattern extraction. Both pipelines implement a two-stage process and leverage an LLM at different stages. The two stages implement \textit{Preprocessing} (P), to select relevant text from the unstructured input report, and \textit{Extraction} (E), to select and classify the target entities. 

The LLM is provided as-a-Service, through API access. While different providers are in principle possible (including self-hosting), aCTIon currently supports the entire GPT family from OpenAI~\cite{openai-api}. In this paper we specifically focus on the \texttt{gpt-3.5-turbo} model~\cite{openai-gpt3.5}.

Finally, \textit{Data Exporter} parses the output of the pipelines to generate the desired output format, i.e., STIX bundles.

\vspace{0.2cm}
\noindent\textbf{Design challenges and decisions}
The aCTIon's two-stages pipelines are designed to handle the two main challenges we faced during the design phase.
First, our main concern has been related to the handling of LLM's hallucinations, such as made-up malware names.
To minimize the probability of such occurrences, we decided to use the LLM as a \textit{reasoner}, rather than relying on its \textit{retrieval} capability. That is, we instruct the LLM to use only information that is exclusively contained in the provided input. For instance, we always provide the definition for an entity we want to extract, even if the LLM has in principle acquired knowledge about such entity definition during its training.
Nonetheless, this approach relies exclusively on \textit{prompt engineering}~\cite{liu2023pre}, and by no means it provides strong guarantees about the produced output. Therefore, we always introduce additional steps with the aim of verifying LLM's answers. These steps might be of various types, including a second interaction with the LLM to perform a self-check activity: the LLM is prompted with a different request about the same task, with the objective of verifying consistency. Finally, we keep CTI analysts in the output verification loop, always including in our procedures the STIX bundle review step as described in Section~\ref{sec:dataset:methodology}.

A second challenge is related to the input size limitations. Our current target model supports a maximum of 4k tokens to be shared in between input and output size. This budget of tokens has to suffice for: (i) the instruction prompt; (ii) any definition, such as what is a malware entity; (iii) the entire unstructured input text; (iv) the produced output. Taking into account that reports in our dataset can be over 6k words long, we had to introduce ways to distill information from the unstructured text, before performing information extraction. Our solution was to introduce the pre-processing steps in our pipelines, with the purpose of filtering, summarizing and selecting text from unstructured inputs. Like in the case of the self-check activity, we might leverage the LLM to perform text selection and summarization.  

\subsection{Entity and Relation Extraction Pipeline}
For the entity and relations pipeline, the preprocessing step performs iterative \textit{summarization}~\cite{wu2021recursively}.
First, the input text is split in chunks of multiple sentences, then each chunk is summarized using the LLM, with the following prompt.
\begin{tcolorbox}[colback=red!5!white,colframe=red!100!black,title=Preprocessing prompt,fontupper=\footnotesize,fontlower=\footnotesize,boxsep=2pt,left=2pt,right=2pt,top=2pt,bottom=2pt]
\texttt{Write a concise summary of the following:\\   
\{text\}\\
CONCISE SUMMARY:}
\end{tcolorbox}
The generated summaries are joined together in a new text that is small enough to fit in the LLM input. This process could be repeated iteratively, however in our experience a single iteration is generally sufficient.

The extraction stage takes as input the summarized report and performs as many requests as entities/relations that need to be extracted. Each provided prompt contains: (i) a definition of the entity that should be extracted; (ii) a direct question naming such entity. A (partial) example follows.

\begin{tcolorbox}[colback=green!5!white,colframe=mygreen!100!black,title=Entity Extraction prompt,fontupper=\footnotesize,fontlower=\footnotesizeboxsep=2pt,left=2pt,right=2pt,top=2pt,bottom=2pt]
\texttt{Use the following pieces of context to answer the question at the end. \\
\{context\}\\
Question: Who/which is the target of the described attack?}
\end{tcolorbox}

After the extraction, the pipeline performs a check-step, querying the LLM to confirm that the extracted entity/relation is present in the original text, and reporting an error in case of inconsistency. In our tests no errors were reported. 

\subsection{Attack Pattern Extraction Pipeline}
\label{sec:action:pipeline}
Extracting Attack Patterns differs significantly from the extraction of simpler entities. As introduced earlier, this task is about identifying behaviors described in the text and associating them to a definition of a behavior according to the MITRE ATT\&CK Matrix taxonomy. Given the large number of attack patterns in the MITRE ATT\&CK Matrix, it would be inefficient (and expensive) to query the LLM directly for each attack pattern's definition and group of sentences in the input report.\footnote{Assuming that a report has 10 paragraphs to inspect, and considering the over 400 techniques described the MITRE ATT\&CK Matrix, we would need to perform over 4k LLM requests for a single report!} We rely instead on an approach that is in principle similar to previous work~\cite{ladder-github, tram-github}: we check the similarity between embeddings of the report's sentences and of the attack pattern's description examples. An embedding is the encoding of a sentence generated by a language model\footnote{aCTIon computes the embeddings using \texttt{text-embedding-ada-002} model from OpenAI \cite{openai-text-embeddings}, but our approach is generic and not tied to a specific text embeddings generation method.}. The language model is trained in such a way that sentences with similar meanings have embeddings that are \textit{close} according to some distance metric (e.g., cosine similarity). Thus, our pipeline's extraction stage compares the similarity between report sentences embeddings and the embeddings generated for the attack pattern examples provided by MITRE.

However, differently from the state-of-the-art, we design a pre-processing stage with the goal of generating different descriptions for the same potential attack patterns contained in the report. Here, an important observation is that an attack pattern might be described in very heterogeneous ways. Therefore, for the preprocessing our goal is to generate multiple descriptions of the same attack pattern, to enhance our ability to discover similarities between such descriptions and the taxonomy's examples. In particular, we introduce three different description generation strategies.

The first strategy prompts the LLM to extract blocks of raw text or sentences that explicitly contain formal descriptions of attack patterns. The output of this strategy is generally a paragraph, or in some cases a single sentence. The example prompt follows.

\begin{tcolorbox}[colback=red!5!white,colframe=red!65!black,title=Attack Pattern Extraction preprocessing strategy \#1,fontupper=\footnotesize,fontlower=\footnotesizeboxsep=2pt,left=2pt,right=2pt,top=2pt,bottom=2pt]
\texttt{Use the following portion of a long document to see if any of the text is relevant to answer the question. 
Return any relevant text verbatim.\\
\{text\}\\
Question: Which techniques are used by the attacker? \\
Report only Relevant text, if any}
\end{tcolorbox}

The second strategy leverages the LLM's reasoning abilities and prompts it to describe step-by-step the attack's events, seeking to identify implicit descriptions~\cite{kojima2022large}. The output of the second strategy is a paragraph.

\begin{tcolorbox}[colback=red!5!white,colframe=red!65!black,title=Attack Pattern Extraction preprocessing strategy \#2,fontupper=\footnotesize,fontlower=\footnotesizeboxsep=2pt,left=2pt,right=2pt,top=2pt,bottom=2pt]
\texttt{Describe step by step the key facts in the following text:\\
\{text\}\\
KEY FACTS:}
\end{tcolorbox}

Finally, the third strategy simply applies sentence splitting rules on the input text, similarly to what happens in previous work, and provides single sentences as output.

All the outputs of the three selection strategies are passed to the extraction step, where they are individually checked for similarity with the MITRE taxonomy's examples. We empirically define similarity threshold, after which we assign to the examined text block the attack pattern's classification of the corresponding MITRE taxonomy's example.

The three strategies extract complementary and non-overlapping information. We report a more detailed analysis in Section \ref{sec:eval:ape}.

\section{Evaluation}
\label{sec:eval}
The need to reduce the effort of CTI analysts in our organization pushed us to test many solutions and previous works. In this section, we present the evaluation of aCTIon in comparison to such solutions, which we consider as performance baselines. We start presenting the implementation of the baselines (which were not always available in open source) and the evaluation metrics, to then present results on the dataset introduced in Section~\ref{sec:dataset} and an ablation study.

\subsection{Baselines Selection and Implementation}
\label{sec:baselines}

For baselines implementation, we followed the following principles. 
First, we aimed to include at least one representative method for each family of NLP algorithms used in the literature. 
Second, whenever possible, we used the original open-source implementation of the methods. When this was not possible, we relied on the original implementation of the underlying NLP algorithm.
Third, we trained or fine-tuned all methods using the same dataset (when possible). 
In fact, the NLP-based methods we tested can be leveraged in two different ways. One approach is to train them on general data and (later referred as domain-agnostic models) use them directly. Another approach is to further fine-tune them on CTI-specific data, before using them. 
Finally, we used the default hyperparameters as described in the corresponding papers.
We discuss the implemented solutions next, dividing them among solutions that focus on general entity and relations extraction, and solutions that deal specifically with attack pattern extraction.

\textbf{Entity and Relations Extraction}
As explained in Section~\ref{sec:background:automation}, a Named Entity Recognition (NER) solution is the basic building block of previous work. The three main families of models used in state-of-the-art NER tasks are Convolutional Neural Networks (CNN), BiLSTM \cite{ma2016end} and Transformers \cite{vaswani2017attention}. 
Among the previous work targeting structured CTI extraction, GRN \cite{chen2019grn} relies on CNN, ThreatKG \cite{gao2022threatkg} and CASIE \cite{satyapanich2020casie} are based on BiLSTM, FLERT \cite{schweter2020flert} and LADDER \cite{alam2022looking} are based on Transformers.
In the case of CNN and BiLSTM methods, models are specifically trained end-to-end for the Entity extraction task. 
Instead, approaches based on Transformers typically rely on pre-trained language models that are trained either on a general-domain corpus or a corpus that includes both general and domain-specific documents~\cite{aghaei2023securebert}. These models are then fine-tuned on a labeled dataset for the Entity extraction task.
For all approaches, a word-level CTI dataset is required. This dataset consists of a CTI corpus where individual words in a sentence have been annotated with tags that identify the named entities according to a specific CTI ontology and a specific format, such as the BIO format \cite{ramshaw1999text}.

The labeling for this task is complex and time consuming~\cite{park2022full, lim2017malwaretextdb}, moreover it requires cross-domain expertise. In fact, CTI experts need to be also familiar with NLP annotation techniques, which makes generating such datasets challenging. Thus, we couldn't use our dataset as training corpus, and instead we rely on a publicly accessible dataset.

We trained all the selected models on the same dataset, using the same train/test/validation set split, to ensure a fair comparison. Specifically, we used the word-level dataset provided in \cite{ladder-github}, which has also been used in previous works such as \cite{alam2022looking}. We chose this dataset because it is the largest open-source dataset available and it is labeled according to an ontology that can be easily mapped to the STIX.
We then test the trained methods and tools performance using our dataset, since it focuses on CTI-metrics.

For CNN-based NER, we use the original open-source implementation~\cite{grn-github} of GRN~\cite{chen2019grn}.
For BiLSTM-based models, we use the domain-agnostic open-source implementation~\cite{bilstm-github} from \cite{reimers2017reporting}. Indeed, ThreatKG~\cite{gao2022threatkg} does not provide an open-source version of their models, and while CASIE~\cite{satyapanich2020casie} does provide an open-source implementation, it cannot be directly adapted to the dataset used to train the other models. Also, their dataset is labeled according to an ontology that is very different from STIX and thus cannot be used for a fair comparison. 
Finally, for Transformer-based models, we present two baselines: one is the original implementation of LADDER~\cite{alam2022looking} as a domain-specific tool, and the other one is a domain-agnostic NER implementation based on FLERT~\cite{schweter2020flert} using the open-source implementation provided in \cite{flair-github, akbik2019flair}.

\textbf{Attack Pattern Extraction} For the Attack Pattern Extraction task, we evaluated a wide range of approaches, namely template matching (TTPDrill \cite{husari2017ttpdrill}, AttacKG \cite{li2022attackg}), Machine Learning (rcATT \cite{legoy2020automated}, TRAM \cite{tram-github}), LSTM (cybr2vec LSTM \cite{padia2018umbc,orbinato2022automatic}), and Transformers (LADDER \cite{alam2022looking}, SecBERT \cite{orbinato2022automatic}). All the baselines target the identification of the attack patterns in terms of MITRE techniques\footnote{All the baselines focus on the main MITRE enterprise techniques without considering the lower-level sub-techniques.} and provided either a pre-trained model or their own dataset to perform the training.

All the methods employ datasets based on the same taxonomy (i.e., MITRE ATT\&CK) and that were directly extracted from the same source, either the description of the MITRE attack patterns or samples of MITRE attack pattern description (both provided by MITRE). Given the high similarity of the datasets in this case, we trained each model using their own dataset.
All the methods were evaluated using their original open-source implementations \cite{ttpdrill-github,attackg-github,rcatt-github,tram-github,secbert-github,ladder-github}.

\subsection{Performance metrics}

We compared each method against the Ground Truth (GT) from our dataset using the following metrics as defined in \cite{orbinato2022automatic}:
\begin{itemize}
    \item Recall: fraction of unique entities in the GT that have been correctly extracted
    \item Precision: fraction of unique extracted entities that are correct (i.e. part of the GT)
    \item F1-score: harmonic mean of Precision and Recall
\end{itemize}
For the sake of clarity, in the rest of this section we use "entities" to refer to both entities and attack patterns, i.e. the outcomes of the two extraction tasks.
From a high-level perspective, the Recall indicates, for a given report, how much the GT entities have been \textit{covered} by a method.
The Precision is instead impacted by the extracted entities which are wrong (i.e. False Positive), e.g. the ones extracted with a wrong type or the ones that the human annotator has not selected as relevant enough.
On the contrary, a True Positive refers to an entity that has been correctly identified with the proper type and with the same text as in the GT.
Finally, a False Negative refers to an entity present in the GT but that has been \textit{missed} by the method at hand.

A na\"ive tool extracting \textit{all} the possible entities of a given type might have a very high Recall but a very low Precision.
A good tool should rather balance both metrics, especially when used to help the annotation task of the human operator that would otherwise spend a lot of time in checking results with many False Positives.
To further investigate this aspect, we also provide the number of entities reported by each tool and we compare it to the numbers from the GT.
We perform this investigation only for the Attack Pattern Extraction task because, based on a simple analysis on the GT, there are an order of magnitude more Attack Patterns than the other types of entities, making this issue particularly important.

In the following subsections, we compute these metrics for \texttt{Malware}, \texttt{Threat Actor}, \texttt{Identity} pointed by a \texttt{targets} relation (which we call just \texttt{Target}) and \texttt{Attack Pattern}.
We adopt the same methodology of \cite{orbinato2022automatic}, compute the metrics per each report, and then provide aggregate statistics.
Given the nature of the GT (with some reports having just 0 or 1 entity of a given type), some metrics exhibit a bimodal distribution across the reports, i.e. they can be either 0 or 1.
In order to provide better visibility of the underlying distribution of the values, we selected violin plots \cite{hintze1998violin} in place of boxplots.
Still, to show at-a-glance the data range, we also report both the average across reports (with a marker) and the 25th and 75th percentiles (as vertical bars).

\subsection{Entity Extraction}

\begin{figure}[t]
\centering{\includegraphics[width=\columnwidth]{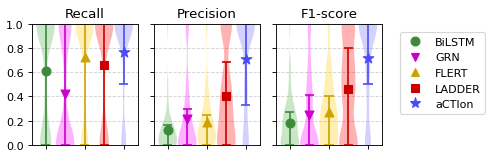}}
\caption{Malware Entity extraction: aCTIon shows better performance compared to the baselines, with an improvement of over 25\%points in F1-score compared to the best performing baseline.}
\label{fig:ner_malware}
\end{figure}

\begin{figure}[t]
\centering{\includegraphics[width=\columnwidth]{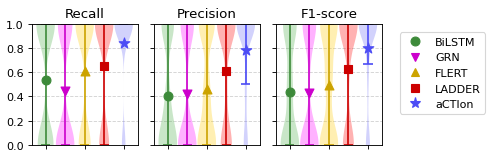}}

\caption{Threat Actor Entity extraction: aCTIon shows better performance compared to the baselines, with improvement of over 20\%points in F1-score compared to the best performing baseline.}
\label{fig:ner_threat_actor}
\end{figure}

Figures \ref{fig:ner_malware} and \ref{fig:ner_threat_actor} show the results for aCTIon against the baselines for Malware and Threat Actor entities, respectively.
aCTIon outperforms the other baselines in terms of Recall, Precision, and consequently F1-score, for both entity extraction tasks.
aCTIon achieves an average Recall, Precision and F1-score of 77\%, 71\% and 72\%, respectively, for the Malware entity extraction and 84\%, 78\% and 80\% for the extraction of Threat Actor entities. This is an increment of over 25\%points for Malware, and about 20\%points for Threat Actor when comparing the F1-score with the best performing baseline (LADDER).
To explain this performance difference, we inspected where the baselines fail, and identified two main cases: (i) baselines fail to understand when an entity is not relevant for the target STIX bundle; (ii) they tend to wrongly include entities that are conceptually close to the entity of interest (e.g., they select a named software in place of a Malware).

For example, when considering our example report from Section~\ref{sec:background}, all baselines can identify HelloXD as malware, resulting in the same recall performance (for this specific report) to that achieved by aCTIon. However, the precision is much lower. This is because baseline methods include entities such as Lockbit 2.0 and x4k (the name of the threat actor and their several aliases) among the detected malwares. Furthermore, they also include a wide range of legitimate applications such as Tox, UPX, Youtube, and ClamAV. For instance in the following extract, in contrast to aCTIon, baselines identify ClamAV as a malware:
\quotebox{
led us to believe the ransomware developer may like using the ClamAV branding for their ransomware.
}

\begin{figure}[t]
\centering{\includegraphics[width=0.8\columnwidth]{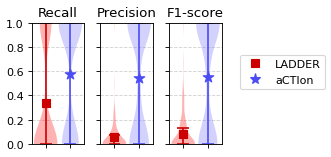}}
\caption{Entity extraction: Target. Only LADDER and aCTIon support the extraction of this type of entity, which involves the assessment of relationships among entities.}
\label{fig:ner_target}
\end{figure}

Figure \ref{fig:ner_target} shows the results for extracting Target entities. This type of entity is trickier, since it requires to understand both named entities, and then the relation among them. Consider the following extract:
\quotebox{HelloXD is a ransomware family performing double extortion attacks that surfaced in November 2021. During our research, we observed multiple variants impacting Windows and Linux systems.}

The above sentences describe the targets of the attack: "Windows and Linux systems". In STIX, they will be classified with the generic Identity class, which includes classes of individuals, organizations, systems, or groups. To correctly identify the Target entity, it is then necessary to understand if the Identity node is an object of a "targets" relation.

Among the considered baselines, only LADDER is capable of extracting this type of entity, as it is equipped with a Relation Extraction model in addition to NER components. Despite the complexity of this task, aCTIon demonstrates its effectiveness by significantly outperforming LADDER also in this case (about 50\%points higher F1-score).

\subsection{Attack Pattern Extraction}
\label{sec:eval:ape}

\begin{figure*}[ht!]
\begin{minipage}[t]{\linewidth}
\includegraphics[width=\linewidth]{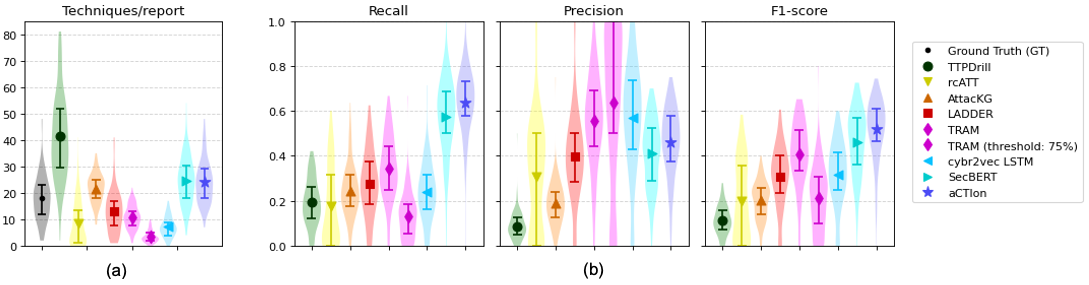}    
 \caption{Attack Pattern extraction. aCTIon outperforms other methods on recall, while achieving a good precision, thereby obtaining about 10\%pts improvement on the F1-score compared to the best performing baseline.}
 \label{fig:atp_eval}
\end{minipage}%
\end{figure*}

For the Attack Pattern Extraction, we focused on a subset of 127 reports that do not include in the text a list or a table of MITRE Techniques in the well-defined \texttt{Txxx} format\footnote{E.g. \texttt{T1573} refers to the MITRE Technique \textit{Encrypted Channel}. The full taxonomy is provided at \url{https://attack.mitre.org/techniques/enterprise/}.}.
Attack Patterns reported with such well-formed format can be trivially extracted with a regex-based approach, and could therefore be trivially detected.

The first plot of Figure \ref{fig:atp_eval}a reports the number of attack pattern extracted from each report by different methods (please notice that, apart from LADDER, the baselines are different from those used for the previous task, cfr. Sec. \ref{sec:baselines}).
The plots of Figure \ref{fig:atp_eval}b report instead the Recall, Precision and F1-score performance metrics.
From a high-level perspective we can divide the baseline methods in two groups.
The first group includes "conservative" methods, i.e., those that tend to limit the number of Attack patterns extracted from each report, resulting in a lower average number compared to the Ground Truth\footnote{The first bar in the violin plot in Figure \ref{fig:atp_eval}a indicates the actual number of techniques per report.}, namely  rcATT, LADDER, TRAM and cybr2vec LSTM. This group is characterized by recall values that are significantly lower than those of precision. 
The second group includes instead methods that have an average number of extracted attack patterns higher than the Ground Truth, and with recall similar to precision, namely TTPDrill, AttacKG and SecBERT.
In the first group, TRAM offers the best performance (F1-score), while in the second group, SecBERT offers the overall best performance.

In addition to being the best within their respective group, these two methods represent two different approaches to the Attack pattern extraction.
Indeed, the former, TRAM, is a framework designed to assist an expert in manual classification, i.e., its output must be reviewed by an expert, so it is important to have high precision and keep the number of attack patterns to be verified low.
This can be achieved, e.g., by increasing TRAM's minimum confidence score from 25\% (its default value) to 75\%.
On the other hand, the latter, SecBERT, is designed to be fully automated.

aCTIon outperforms all the baselines in terms of overall performance (F1-score) by about 10\% point. 
More importantly, the recall is higher than any other solution, and the average precision is about 50\%. These results make a manual verification by CTI analysts manageable: the average number of attack patterns extracted per-report is 25 (cf. Figure~\ref{fig:atp_eval}a).

\noindent
\textbf{Comparison with previous work}
The main difference between aCTIon and state-of-the-art methods is in how they select text that may contain attack patterns. Some methods, such as rcACT, use the entire document, while others (SecBERT and TRAM) use all the sentences in the document. Yet others (such as LADDER and TTPDrill) select a specific subset of sentences. The selected pieces of text are then used as input to a classifier. The advantage of aCTIon is that it can leverage the LLM reasoning capability to improve the selection of text blocks before the classification/extraction.

The following is a paragraph describing the use of attack pattern T1573 \textit{"Encrypted Channel"}\footnote{\url{https://attack.mitre.org/techniques/T1573/}.} extracted from a report: 

 \quotebox{
 The January 2022 version of PlugX malware utilizes RC4 encryption along with a hardcoded key that is built dynamically. For communications, the data is compressed then encrypted before sending to the command and control (C2) server and the same process in reverse is implemented for data received from the C2 server. Below shows the RC4 key "sV!e@T\#L\$PH\%" as it is being passed along with the encrypted data. The data is compressed and decompressed via LZNT1 and RtlDecompressBuffer. During the January 2022 campaigns, the delivered PlugX malware samples communicated with the C2 server 92.118.188[.]78 over port 187.
 }

How PlugX leverages RC4 encryption to communicate with the command and control, and thus that is an obfuscated communication channel is clear only after a few sentences.
Some of the state-of-the-art methods will miss this attack pattern because they act at sentence level. Some others may miss it because the selection of the sentences depends on a previous correct identification of the involved entities in the description (and in the previous section we show how this step can fail). Finally, for those methods which use the whole document, there is no guarantee that the document-level representation will capture this information.

aCTIon uses two different strategies for selecting the text block. The first strategy prompts the LLM to retrieve portions of the text that contains the attack pattern description (i.e., more than one sentence). However,  the LLM may not recognize the specific attack pattern, indeed there is no guarantee that the LLM knows this specific attack pattern. To avoid such cases, a second strategy is used together with the previous one. The LLM is also prompted to reason about the key steps performed in the attack. The sentence below is the output of the second strategy.

 \quotebox{
The January 2022 version of PlugX malware uses RC4 encryption with a dynamically built key for communications with the command and control (C2) server.
} 

This sentence not only clearly expresses the attack pattern but it easier to process in the classification step. 
Finally, we also process all the sentences in the text separately.
How these strategies contribute to the final result is discussed in the next section.

\subsection{Ablation Study}
\label{sec:eval:ablation}
We conducted an ablation study on the preprocessing step for both the Entity and Attack Pattern extraction pipelines. Indeed, it is crucial to understand how information is selected and filtered in this step. 

\begin{figure}[t]
\centering{\includegraphics[width=0.9\columnwidth]{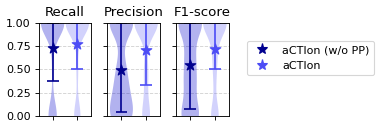}}
\caption{Ablation study: Malware extraction w/o and w/ preprocessing (PP). The preprocessing step does not introduce any loss of information (i.e., comparable recall), while filtering out non-relevant information (i.e. improving the precision).}
\label{fig:mal_ablation}
\end{figure}

\begin{figure}[t]
\centering{\includegraphics[width=0.9\columnwidth]{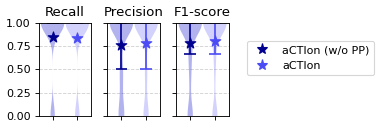}}
\caption{Ablation study: Threat Actor extraction w/o and w/ preprocessing (PP). In this case the performance are similar because few Threat Actors are mentioned in each report and they are typically relevant.}
\label{fig:ta_ablation}
\end{figure}

\begin{figure}[t]
\includegraphics[width=\columnwidth]{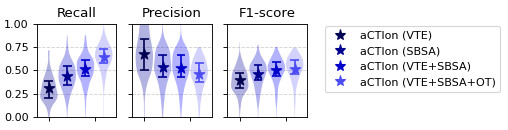}
\caption{Ablation study: Comparison of different preprocessing strategies and configurations for Attack Pattern extraction. aCTIon (VTE) captures explicit attack pattern description with high precision, while aCTIon (SBSA) captures implicit ones. When combined together they show better global performance, confirming they extract non-overlapping, complementary information. }
\label{fig:atp_ablation}
\end{figure}

For Entity extraction, we compared a configuration in which the preprocessing step was included (\textit{aCTIon}) to a configuration where it was omitted (\textit{aCTIon (w/o PP)}).
When omitted, the input report was provided in separate chunks if larger than the available input size.
Figures \ref{fig:mal_ablation} and \ref{fig:ta_ablation} present the performance for the two different configurations when extracting Malware and Threat Actors entities, respectively. The decrease in precision for \textit{aCTIon (w/o PP)} was expected because non-relevant information was still present in the text during processing. Additionally, since only a few Threat Actors are usually present in the same report, the drop in performance was more noticeable for the Malware entity.

It is noteworthy that the preprocessing step does not result in any decrease in Recall, indicating that no relevant information is lost in the summary produced by the LLMs. Additionally, for Malware entities, the recall on preprocessed text is even slightly higher. We conclude that reformulating and summarizing concepts during preprocessing can aid in the extraction process.

In the case of attack pattern extraction, we utilize the preprocessing module to select and enhance text extracts that potentially contain descriptions of attack patterns. The objective of our ablation study is to examine how different preprocessing strategies contribute to identifying such descriptions. We present four variants of our method that correspond to various preprocessing configurations and report their performance in Figure \ref{fig:atp_ablation}.

The first configuration, \textit{aCTIon (VTE)}, selects verbatim text excerpts (strategy \#1 from Section~\ref{sec:action:pipeline}) that may contain an attack pattern. This results in a few attack patterns per report, with high precision (above 67\%). However, it has lower recall, since it is unusual for all attack patterns to be explicitly described in the text.
In the second configuration, \textit{aCTIon (SBSA)} (strategy \#2), the preprocessing is configured to describe the step-by-step actions performed during the attack, aiming to capture implicit or not-obvious descriptions of attack patterns. Using this configuration, we match the global performance (F1-score) of the best state-of-the-art method (SecBERT), while outputting on average 14.4 attack patterns per report - on average, half of what is produced by SecBERT.
The third configuration, \textit{aCTIon (VTE+SBSA)},  uses both preprocessing strategies together, resulting in improved performance. Additionally, it shows that the proposed preprocessing methods extract non-overlapping, complementary information.
Finally, the fourth preprocessing configuration, \textit{aCTIon (VTE+SBSA+OT)}, is our chosen configuration described in Section~\ref{sec:action:pipeline} and labeled \textit{aCTIon} in Figure \ref{fig:atp_eval}. We report it again here for convenience.

\subsection{Multi-language support}
\label{appendix:multi_language}

Not all the CTI sources are in English language.
This is an issue for the analyst that not only should have expertise in the cybersecurity domain, but would also need to know fluently more than one language.
Automated tools can be used only when specialized on multiple languages and this is typically not the case.
Among our baselines, only LADDER includes a multi-language model (XLM-RoBERTa \cite{conneau2019unsupervised}) that can work on non-english reports.
However, this only applies to its NER components for the Entity Extraction task.
Indeed, when considering its 3-stages Attack Pattern Extraction pipeline, all three stages (based on RoBERTa \cite{liu2019roberta} and Sentence-BERT \cite{reimers2019sentence}), only support the English language, making them unsuitable for other languages.
Our method is based on an LLM that during training was exposed to a huge corpus of text including a variety of languages and thus can process reports in languages other than English \cite{bang2023multitask,lai2023chatgpt}.

Out of our dataset, 13 reports are also provided in an alternative language other than English, such as Japanese.
We evaluated the performance of aCTIon against LADDER for the Entity Extraction task and we reported in Table \ref{table:non_eng_reports} the average performance across the 13 reports.
In general, we can observe that aCTIon and LADDER present a gap of performance comparable to the one observable when analyzing the English reports (9\%point-35\%point).
In the case of Attack Pattern Extraction, only aCTIon is able to produce a usable result, and its performances are comparable to what obtained for English reports.

\begin{table}[t]
\caption{Entity extraction on non-English reports}
\small
\begin{center}
\begin{tabular}{|c||c|c|c||c|c|c|}
\hline
             & \multicolumn{3}{c||}{LADDER} & \multicolumn{3}{c|}{aCTIon} \\ \hline 
               & Rec & Prec & F1 & Rec & Prec & F1 \\ \hline
\hline
Malware        & 0.73 & 0.58 & 0.61 & 0.92 & 0.67 & 0.75 \\ \hline
Threat Actor   & 0.46 & 0.42 & 0.44 & 0.69 & 0.65 & 0.67 \\ \hline
Victim         & 0.31 & 0.19 & 0.21 & 0.54 & 0.54 & 0.54 \\ \hline
Attack Pattern & 0.01 & 0.08 & 0.01 & 0.44 & 0.52 & 0.46 \\ \hline
\end{tabular}
\end{center}
\label{table:non_eng_reports}
\end{table}

\section{Discussion}
\label{sec:discussion}
\noindent\textbf{Limitations of the Evaluation} 
Our evaluation focused on \texttt{Malware}, \texttt{Threat Actor}, \texttt{Target} and \texttt{Attack Pattern} entities. This was the case because these entities enabled us to directly compare with previous work, i.e., other entities were not widely supported by other tools. As a result, our evaluation did not extensively cover the extraction of relations. In fact, only the extraction of \texttt{Target} includes relation extraction (of type \texttt{targets}), and we could only compare to LADDER that supports it (cf. Figure~\ref{fig:ner_target}). However, aCTIon can extract any relation defined in the STIX's ontology, and therefore we assessed the performance also in that regard. For example, for all relations between \texttt{Malware}, \texttt{Threat Actor} and \texttt{Identity}, i.e., relation of types \texttt{targets} and \texttt{uses}, aCTIon achieves 73\% recall and 88\% precision on average (cf. Appendix~\ref{appendix:relation}). 

\noindent\textbf{Deployment advantages}
We focused this paper on performance results, however in a practical setting it is important to consider ease of \textit{deployment} and \textit{maintenance} among the goals. Compared to previous work, the reliance of aCTIon on LLMs removes the need to collect, annotate and maintain datasets for the different components of the previous work's pipelines (e.g., NER components). Furthermore, previous work, e.g., LADDER, makes extensive use of hand-crafted heuristics to clean and filter the classification output (e.g., count-based filters to remove noisy entities and ambiguities, or allow-lists of know non-malicious software). Heuristics also require continuous maintenance and adaptation. In contrast, aCTIon does not require the collection and annotation of datasets, nor the use of hand-crafted heuristics (cf. Appendix~\ref{appendix:ner_baselines}).
It is important to note that heuristics are data-dependent, meaning that a change in the dataset's characteristics might require a corresponding change in the utilized heuristics (this is the case, e.g., for the count-based heuristics of LADDER). On the other hand, the prompts used in the aCTIon pipeline are solely task-dependent.

\noindent\textbf{Is the problem solved?} 
Arguably, most of the benefits of aCTIon are derived from the proper use and integration of LLMs in the information extraction pipeline. Our tests with this technology were initially purely exploratory, but as we employed it increasingly in testing deployments, we grew confidence that tools like aCTIon can already do most of the heavy lifting in place of CTI analysts, for structured CTI extraction. We expect performance will continue to improve with the development of more powerful LLMs (e.g., GPT4 was released during the writing of this paper) that allow for larger input sizes and better reasoning capabilities. Therefore, we expect the recall and precision metrics to further improve without significant changes to aCTIon.
Nonetheless, we also inherit the shortcomings of LLMs, such as \textit{hallucinations}~\cite{lee2022factuality,ji2023survey}. 
CTI analysts are required to carefully review the outputs of the system. 
This issue is what currently hinders the full automation of the structured CTI information extraction tasks.

\hfill\\\textbf{Hallucinations.}  One well-known challenge of LLMs is their tendency to produce nonfactual, untruthful information, which is commonly referred to as "hallucinations" \cite{lee2022factuality,ji2023survey}. To mitigate this issue, aCTIon has implemented several measures. Firstly, the LLM is instructed to rely solely on the input CTI report provided to it, without considering any additional knowledge acquired during training. Secondly, for entity extraction, it is possible to verify if the model circumvented this limitation by checking if the extracted entities are actually present in the text. During our tests, we only found 2 (0.9\%) cases of this for both Malware and Threat actors. Additionally, aCTIon uses only user-provided classes to label entities and attack patterns, thus the model cannot hallucinate labels as output. Furthermore, any errors in classification will have the same impact as simple misclassification.

\noindent\textbf{Other issues} 
The current precision and recall of aCTIon is within the 60\%-90\% range for most entities. In our experience, this is already in line with the performance of a CTI analyst, and unlike human analysts, aCTIon keeps consistent performance over time not being affected by tiredness. 
In fact, we speculate that many misclassifications are an artifact of the ambiguous semantics associated with CTI and the related standard ontologies. For example, what is considered a relevant entity by an analyst may differ. Defining clear semantics for CTI data remains an open challenge.

\noindent\textbf{Ethical considerations} 
In this paper we do not collect/use any user data nor perform unauthorized experiments over third party infrastructures. Our dataset only includes already publicly accessible reports about cyber threats, shared by the security community. We believe that this information can help defending against such threats, and we do not foresee any harmful use of the shared structured information that would out-weight the benefits. 
We further verified each entry to exclude any potentially confidential or sensible information.
Reports are shared with a link to the original source. Some reports and STIX bundles in the dataset may contain explicit Indicators of Compromise (IoCs). We have not redacted them from the public version of the dataset as we believe that this information may be valuable for future research. Some reports and their respective STIX objects may include offensive language such as malware or threat actor names. We have not redacted this information because it would affect the usefulness of the data. Lastly, reports and their respective STIX bundles may attribute attacks to specific groups of people or states. When considering such information, the full report text should be considered to put in context why the attribution was suggested by their authors.
Finally, regarding aCTIon, all the STIX bundles it automatically extracts are always verified by CTI experts. 

\section{Related work}
\label{sec:related}
We covered related work on LLMs in Section~\ref{sec:action} and on structured CTI extraction in Section~\ref{sec:eval}, therefore here we focus on works that leverage the structured information. In fact, we already mentioned that structured CTI enables the investigation and monitoring activities inside an organization, i.e. threat hunting, based for example on TTPs \cite{daszczyszak2019ttp,sharma2023ttp, gao2021system}. However, there are also other relevant uses.
In particular, trend analysis and prediction of threats for proactive defense take advantage of structured CTI~\cite{chierzi2021evolution,alam2022looking,oosthoek2019sok,al2020learning}. Adversary emulation tools like MITRE CALDERA \cite{applebaum2016intelligent,mitre-caldera-github} can also benefit from structured CTI because they are typically fed with adversarial techniques based on e.g., MITRE ATT\&CK.

\section{Conclusion}
We introduced a dataset to benchmark the task of extracting structured Cyber Threat Intelligence (CTI) from unstructured text, and aCTIon, a solution to automate the task. 

The dataset is the outcome of months of work from our CTI analysts, and provides a structured STIX bundle for each of the 204 reports included. We release it openly. To the best of our knowledge, the dataset is 34x larger than any other publicly available dataset for structured CTI extraction, and the only one to provide complete STIX bundles.

We then introduced aCTIon, a framework that leverages recent advances in Large Language Models (LLMs) to automate structured CTI extraction. To evaluate aCTIon we selected 10 different tools from the state-of-the-art and previous work, re-implementing them when open source implementations were not available.
Our evaluation on the proposed benchmark dataset shows that aCTIon largely outperforms previous solutions.
Currently, aCTIon is in testing within our organization for daily production deployment.

There has been a recent wave of announcements of security products based on LLMs to analyze and process CTI~\cite{MicrosoftSecurityCopilot, RecordedFuture, GoogleSecurityWorkspace,CrowdStrikeCharlotteAI,SentinelOne}. However, the security community was still lacking a benchmark that would allow the evaluation of such tools on specific CTI analysts tasks. Furthermore, there is a lack of information about how LLMs could be leveraged in this area for the design of systems. 
We believe our work provides both a way to benchmark such new tools, with our dataset, and a first system design and set of insights to leverage LLMs in CTI tasks.





%

\bibliographystyle{IEEEtran}  
\bibliography{biblio}

\clearpage

\appendices


\section{Relationship extraction}
\label{appendix:relation}

\begin{figure}[t]
\centering{\includegraphics[width=0.8\columnwidth]{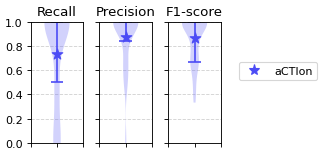}}
\caption{Relation extraction microbenchmark}
\label{fig:rel_extraction}
\end{figure}

Evaluating relationship extraction task is more complex than regular entity extraction. In fact, there is a need to include negative samples of relationships in order to verify that the classifier is able not only to confirm the existence of a relation among entities, but also its lack. Generating random non-existent syntactically correct links is an approach commonly adopted in the state-of-the-art~\cite{yang2014embedding}. 

Thus, we defined a benchmark to evaluate aCTIon performance in extracting STIX relations between entities. 
We use as positive samples (i.e. existing relations between existing entities) all the relations between Malware, Threat Actor, and Identity entities that were present in the STIX representation of the report (i.e. as they were extracted by the CTI analyst).
We use as negative samples (i.e. not existing relations between existing entities) a set of randomly generated relations between entities that are present in the text.
This set may also include entities which are extracted by the Entity extraction task but that have been then filtered out in the final STIX representation by the CTI analyst (e.g. Lockbit 2.0 in the HelloXD report). 
Positive samples and negative samples form our evaluation dataset. 

Furthermore, since the scope of the test is to benchmark just the relation extraction capabilities, we do not use entities extracted by aCTIon (that would automatically filter out some negative samples) but we build this dataset directly from the STIX representation provided by the CTI analyst.

aCTIon is configured to preprocess the text using the same compression techniques described in Section~\ref{sec:action} and is prompted to extract the relation between two entities using a direct question. Figure \ref{fig:rel_extraction} shows the results for all relations between Malware, Threat Actor, and Identity entities, i.e., relations of types targets and uses. On average, aCTIon achieves 73\% recall, 88\% precision and 86\% f1-score.

\section{Baselines with heuristics}
\label{appendix:ner_baselines}

\begin{figure}[t]
\centering{\includegraphics[width=\columnwidth]{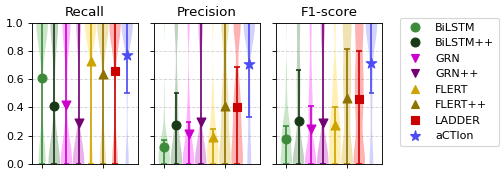}}
\caption{Entity extraction: Malware}
\label{fig:ner_malware_with_heuristics}
\end{figure}

\begin{figure}[t]
\centering{\includegraphics[width=\columnwidth]{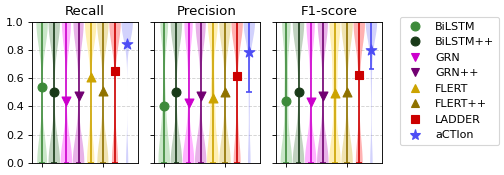}}
\caption{Entity extraction: Threat Actor}
\label{fig:ner_threat_actor_with_heuristics}
\end{figure}

In this section we consider 3 additional baselines where we apply the same post-processing heuristics provided by LADDER to each one of the other 3 baselines, i.e. BiLSTM, GRN and FLERT.
We named these new baselines BiLSTM\texttt{++}, GRN\texttt{++} and FLERT\texttt{++}, respectively.
The post-processing heuristics were used by LADDER to remove noisy entities and solve some ambiguities produced by the NER module.
As shown in Figures \ref{fig:ner_malware_with_heuristics} and \ref{fig:ner_threat_actor_with_heuristics}, the heuristics are able to increase the Precision because some irrelevant entities are filtered out.
At the same time, however, the Recall decreases, implying that the heuristics are also wrongly cutting out some information which is relevant. 
Remarkably, when extracting Malware entities, FLERT\texttt{++} is able to reach the F1-score performance of LADDER, which however still keeps its role of best performing baseline for both types of entity.

\end{document}